\documentstyle[sprocl]{article}

\def\be{\begin{equation}}
\def\ee{\end{equation}}
\def\bea{\begin{eqnarray}}
\def\eea{\end{eqnarray}}

\begin{document}


\title{ON THE WIGNER-RACAH ALGEBRA OF THE GROUP $\mbox{SU}_{2}$ 
IN A NON-STANDARD BASIS} 

\author{ M.R. KIBLER }

\address{Institut de Physique Nucl\'eaire de Lyon\\
IN2P3-CNRS et Universit\'e Claude Bernard\\
43, Boulevard du 11 Novembre 1918\\
F-69622 Villeurbanne Cedex, France}


\maketitle\abstracts{The algebra su$_{2}$ is derived from two commuting quon
algebras for which the para\-me\-ter $q$ is a root of unity. 
This leads to a polar decomposition of the shift operators $J_+$ and
$J_-$ of the group SU$_{2}$ (with $J_+ = J_-^{\dagger} = H U_r$ where $H$ is
Hermitean and $U_r$ unitary). The Wigner-Racah algebra of  SU$_{2}$ is 
developed in a new basis arising from the simultanenous diagonalization of the 
commuting operators $J^2$ and $U_r$.}   

\vskip 6 true cm 

\noindent
Paper based on a lecture given to the Vth 
International
School on Theoretical Physics ``Symmetry and Structural Properties of Condensed
Matter'' (Zaj\c aczkowo, Poland, 27 August - 2 September 1998). To be  
published in {\em Symmetry and Structural Properties of
Condensed Matter}, eds. T. Lulek, B. Lulek and A. Wal (World Scientific,
Singapore, 1999). 

\vfill\eject

\title{ON THE WIGNER-RACAH ALGEBRA OF THE GROUP $\mbox{SU}_{2}$ 
IN A NON-STANDARD BASIS} 

\author{ M.R. KIBLER }

\address{Institut de Physique Nucl\'eaire de Lyon\\
IN2P3-CNRS et Universit\'e Claude Bernard\\
43, Boulevard du 11 Novembre 1918\\
F-69622 Villeurbanne Cedex, France}


\maketitle\abstracts{The algebra su$_{2}$ is derived from two commuting quon
algebras for which the para\-me\-ter $q$ is a root of unity. 
This leads to a polar decomposition of the shift operators $J_+$ and
$J_-$ of the group SU$_{2}$ (with $J_+ = J_-^{\dagger} = H U_r$ where $H$ is
Hermitean and $U_r$ unitary). The Wigner-Racah algebra of  SU$_{2}$ is 
developed in a new basis arising from the simultanenous diagonalization of the 
commuting operators $J^2$ and $U_r$.} 

\section{Introduction}

The concept of a Wigner-Racah algebra (WRa) 
associated to a group takes its origin with
the works by Wigner~$^1$ on a simply reducible 
group (with emphasis on the ordinary rotation group) 
and by Racah~$^2$ on chains of groups of type 
$\mbox{SU}_{ 2 \ell + 1} \supset 
 \mbox{SO}_{ 2 \ell + 1} \supset 
 \mbox{SO}_{ 3 }$ (mainly with $\ell = 2, 3$).
From a practical point of view, the WRa of a group deals with the
algebraic relations satisfied by its coupling 
and recoupling coefficients. From a more
theoretical point of view, the WRa of a finite or compact 
group turns out to be the infinite-dimensional 
Lie algebra spanned by the Wigner  unit  operators 
(i.e., the operators whose matrix elements 
are the coupling or Clebsch-Gordan
or Wigner coefficients of the group). 

The WRa of the group $\mbox{SU}_{2}$
is well known. It is generally developed in the standard basis
$\left\{ |j , m \rangle : 2j \in {\bf N}, 
            \ m = - j, - j + 1, \cdots, j \right\}$
arising in the simultaneous diagonalization of the Casimir 
operator $J^2$ and of one generator, say $J_3$, of 
$\mbox{SU}_{2}$. Besides this basis, there exist 
several other bases. Indeed, any change of basis 
of type
$$
| j , \mu \rangle = \sum_{m=-j}^{j} | j , m \rangle \langle jm | j \mu \rangle 
$$
(where the $(2 j + 1)\times(2 j + 1)$ matrix with elements  
$\langle jm | j \mu \rangle$ is an arbitrary unitary matrix) 
defines another acceptable basis for the WRa of 
$\mbox{SU}_{2}$. In this basis, the matrices of the irreducible 
representation classes of $\mbox{SU}_{2}$
take a new form as well as the coupling coefficients 
(and the associated $3 - jm$ symbols).   As a matter of fact, 
the coupling coefficients $(j_1 j_2 m_1 m_2 | j m)$ are simply replaced by 
   \begin{eqnarray*} 
(j_1 j_2 \mu_1 \mu_2 | j \mu) = 
\sum_{m_1=-j_1}^{j_1} 
\sum_{m_2=-j_2}^{j_2} 
\sum_{m  =-j  }^{j  }     (j_1 j_2 m_1 m_2 | j m) \\
\times                    \langle j_1m_1 | j_1 \mu_1 \rangle^* \,
                          \langle j_2m_2 | j_2 \mu_2 \rangle^* \,
                          \langle j  m   | j   \mu   \rangle
   \end{eqnarray*}
when passing from the $\left\{ j m   \right\}$ scheme to the 
                      $\left\{ j \mu \right\}$ scheme while the recoupling 
coefficients (and the associated $6 - j$, $9 - j$, $\cdots$ symbols)
remain invariant. 

The various bases for  $\mbox{SU}_{2}$  may be classified into two types~:
group-subgroup type and non group-subgroup type. The standard basis corresponds
to a group-subgroup type basis associated to the chain of groups 
 $\mbox{SU}_{2} \supset \mbox{U}_1$. Another 
group-subgroup type basis may be obtained by replacing 
 $\mbox{U}_1$ by a finite group $G^*$ (generally the double, i.e., 
spinor group, of a point group $G$ of molecular or crystallographic interest). 
Among the  $\mbox{SU}_{2} \supset G^*$  bases, we may distinguish~: 
the weakly symmetry adapted bases for which the basis vectors are
eigenvectors of $J^2$ and of the projection operators of $G^*$
(e.g., see Refs.~3-10) and 
the strongly symmetry adapted bases for which the basis vectors are
eigenvectors of $J^2$ and of an operator defined in the enveloping algebra 
of  $\mbox{SU}_{2}$ and invariant under the group $G$ (e.g., see Refs.~11-13). 

It is the aim of this lecture to report some preliminary 
results on an apparently 
non group-subgroup type approach to the WRa of $\mbox{SU}_{2}$. 

The school was dedicated to the memory of Giulio Racah. Professor Racah 
got interested at the end of the fiftees with 
 weakly symmetry adapted $\mbox{SU}_{2} \supset G^*$  bases, mainly in
connection with electron paramagnetic resonance of partly-filled shell 
ions in crystals. His ideas on this subject were developed 
by colleagues, students and students of his students (see for instance 
Refs.~3-6). 

\section{A Quon Realization of $\mbox{SU}_{2}$}

\subsection{Two constituent quon algebras} 

Let us consider the two commuting quon algebras 
$A$ and $B$ spanned by the triplets 
$\left( a_{-}, a_{+}, N_a \right)$ and
$\left( b_{-}, b_{+}, N_b \right)$ of linear operators which satisfy
     $$
   a_{-} a_{+} - 
 q a_{+} a_{-} = 1, \quad 
   N_a a_{\pm} - a_{\pm} N_a = \pm a_{\pm}, \quad
   \left( a_{-} \right)^k = 
   \left( a_{+} \right)^k = 0 
       $$
and 
     $$
   b_{-} b_{+} - 
 q b_{+} b_{-} = 1, \quad  
   N_b b_{\pm} - b_{\pm} N_b = \pm b_{\pm}, \quad 
   \left( b_{-} \right)^k = 
   \left( b_{+} \right)^k = 0 
       $$
where  
     $$
  q = \exp \left( {\rm i} {2 \pi  \over k} \right) \quad {\rm with} \quad 
  k \in {\bf N} \setminus \{ 0,1 \} 
       $$
The parameter $q$ is the same for both algebras 
so that the two algebras $A$ and $B$ are indeed two copies of the same algebra. 
The important difference between the algebra $A$ (or $B$) and the quon algebra
introduced by Arik and Coon~$^{14}$ is to be found in the fact that the
deformation parameter $q$ is here a root of unity instead of being a (positive)
real number. As a consequence, the operators $a_{+}$ and $b_{+}$ are not the
adjoints of                    the operators $a_{-}$ and $b_{-}$, respectively. 
Furthermore, the operators $N_a$ and $N_b$ are supposed to be Hermitean 
operators as in the case where $q \in {\bf R}^*_{+}$ (which includes the
non-deformed case $q=1$). 

The nilpotency conditions 
     $$
   \left( a_{-} \right)^k = 
   \left( a_{+} \right)^k =
   \left( b_{-} \right)^k = 
   \left( b_{+} \right)^k = 0 
       $$
take their origin in the hypothesis that $q^k = 1$. They may be justified 
from the relations 
   $$ 
a_{-} \left( a_{+} \right)^k   = \left( a_{+} \right)^k a_{-},         \quad 
\left( a_{-} \right)^k a_{+}   = a_{+} \left( a_{-} \right)^k                
     $$
and 
   $$
N_a \left( a_{+} \right)^{ k } = \left( a_{+} \right)^{ k } (N_a + k), \quad    
\left( a_{-} \right)^{ k } N_a = (N_a + k) \left( a_{-} \right)^{ k } 
     $$ 
(and similar relations with $a \rightarrow b$) which follow from the 
defining relations 
$a_{-} a_{+} - q a_{+} a_{-} = 1 $ and 
$\left[ N_a , a_{\pm} \right] = \pm a_{\pm}$ 
(and similar relations with $a \rightarrow b$). 

The value $k = 0$ is excluded since it would lead to a non-defined value of
$q$. The case $k = 1$ must be excluded too since it would yield 
 trivial algebras with   
 $a_{-} = a_{+} = b_{-} = b_{+} = 0$. We observe that for $k=2$ (i.e., for 
$q=-1$),
the algebra $A$ (or $B$) corresponds to the ordinary fermionic algebra. On the
other hand, we note that the algebra $A$ (or $B$) corresponds to the ordinary
bosonic algebra in the limiting situation where $k \to \infty$ (i.e., for 
$q = 1$). The algebras $A$ and $B$ with $N_a = N_b$ 
may be used for defining $k$-fermions~$^{15}$ which 
are objects interpolating between fermions and bosons 
like anyons.~$^{16}$

We now define two representations of $A$ and $B$. For the algebra 
$A$, we choose the representation defined by~$^{17}$ 
  \begin{eqnarray*}
   a_{+} | n_a ) &=&                      | n_a +1), \quad a_{+} |k-1) = 0 \\
   a_{-} | n_a ) &=& \left[ n_a \right]_q | n_a -1), \quad a_{-} |0)   = 0 
  \end{eqnarray*}   
and 
     $$
   N_a   | n_a ) = n_a | n_a ) 
       $$
on a unitary space ${\cal F}_a = \{ |  n_a   ) : n_a = 0, 1, \cdots, k-1 \}$ 
of dimension $k$. We use here the notation 
     $$
  \left[ x \right]_q = \frac{1-q^x}{1-q} \quad {\hbox{for}} \quad x \in {\bf R}
       $$
so that 
$$
[n]_q = 1 + q + \cdots + q^{n-1} \quad \mbox{for} \quad n \in N^*
$$
Similarly for the algebra $B$, we take the representation defined by 
  \begin{eqnarray*}
   b_{+} | n_b ) &=& \left[ n_b +1 \right]_q | n_b +1), \quad b_{+} |k-1) = 0 \\
   b_{-} | n_b ) &=&                         | n_b -1), \quad b_{-} |0)   = 0 
  \end{eqnarray*}
and
     $$
   N_b    | n_b ) = n_b | n_b ) 
       $$
on a unitary space ${\cal F}_b = \{ |  n_b   ) : n_b = 0, 1, \cdots, k-1 \}$ 
of dimension $k$. 

From the spaces ${\cal F}_a$ and ${\cal F}_b$, we get the Fock space 
$$
{\cal F} = {\cal F}_a \otimes {\cal F}_b = 
\{ |  n_a ,  n_b  ) = | n_a ) \otimes | n_b ) : 
                  n_a , n_b  = 0, 1, \cdots, k-1 \}
$$
of finite dimension (dim~${\cal F} = k^2$). Let $H$ and $U_r$ be two 
linear operators on ${\cal F}$ defined as~$^{17}$  
     $$
  H = {\sqrt { N_a  \left(  N_b  + 1 \right) }}
       $$
and
     $$
  U_r = \left[  a_{+}  + 
  {\rm exp} \left( {{\rm i}{{\phi}_r \over 2}} \right) {( a_{-} )^{k-1} \over 
  \left[ k-1 \right]_q!} \right]
        \left[  b_{-}  + 
  {\rm exp} \left( {{\rm i}{{\phi}_r \over 2}} \right) {( b_{+} )^{k-1} \over 
  \left[ k-1 \right]_q!} \right]
       $$
where  $\phi_r$  is  an  arbitrary  real  parameter 
and $\left[ k-1 \right]_q!$ is a $q$-deformed factorial defined through 
     $$
  \left[ n \right]_q! = 
  \left[ 1 \right]_q 
  \left[ 2 \right]_q \cdots 
  \left[ n \right]_q \quad {\rm for} \quad n \in {\bf N}^* 
  \quad {\rm and} \quad \left[ 0 \right]_q! = 1  
       $$
It is straightforward to verify that the action of  $U_r$ on 
 ${\cal F}$ is controlled by 
  \begin{eqnarray*}
U_r | k-1, n_b ) &=& {\rm exp} 
\left( {\rm i} \frac{{\phi}_r}{2} \right) | 0     ,  n_b-1) \quad {\hbox{for}} 
                                        \quad n_b \not= 0   \\
U_r | n_a, n_b ) &=& | n_a +1,  n_b -1) \quad {\hbox{for}} 
                                        \quad n_a \not= k-1 
                     \quad {\hbox{and}} \quad n_b \not= 0   \\
U_r | n_a, 0   ) &=& {\rm exp} 
\left( {\rm i} \frac{{\phi}_r}{2} \right) | n_a +1,  k  -1) \quad {\hbox{for}} 
                                        \quad n_a \not= k-1 
  \end{eqnarray*}
and
     $$
  U_r |k-1, 0) = {\rm exp} \left( {{\rm i}{{\phi}_r }} \right) |0, k-1)
       $$
As a consequence, we can prove the identity
     $$
  \left( U_r \right)^{k} = {\rm exp} \left( {\rm i} \phi_r \right) 
       $$  
(the unit matrix of dimension $k^2$ is supposed to occur in the 
right-hand side of the identity). 
The action of $H$ of ${\cal F}$ is much more simple. It is described by 
     $$
  H | n_a ,  n_b ) = {\sqrt{  n_a  ( n_b  + 1) } | n_a ,  n_b )}
       $$
which holds for $n_a = 0, 1, \cdots, k-1$ and
                $n_b = 0, 1, \cdots, k-1$. 

To close this subsection,  it is interesting to note that we can generate the
infinite dimensional Lie algebra $W_{\infty}$ from the generators of the quon
algebras $A$ and $B$. Indeed, by putting 
     $$ 
  U = U_r, \quad V = q^{  N_a  -  N_b  }
       $$
and
     $$
  T_{ (m_1,m_2) } = q^{m_1m_2} U^{m_1} V^{m_2}
       $$
(with $m_1 \in {\bf N}$  and  
      $m_2 \in {\bf N}$) we can prove that~$^{17}$ 
     $$
  \left[ T_m,T_n \right] = -2 \> {\rm i} 
  \sin \left( {2 \pi \over k} m \times n \right) 
  T_{ m+n }
       $$
where we use the abbreviations 
     $$
  m = \left( m_1, m_2 \right), \quad 
  n = \left( n_1, n_2 \right)
       $$
and
     $$
  m+n = \left( m_1+n_1 , m_2+n_2  \right), \quad
  m \times n = m_1 n_2 - m_2 n_1 
       $$
As a result, the operators $T_{\ell}$ span the 
algebra  $W_{\infty}$  introduced  by Fairlie, Fletcher and Zachos.~$^{18}$ 
This result parallels a similar result obtained in Ref.~15 in the study 
of $k$-fermions and of the Dirac quantum phase operator. 

\subsection{A quon approach to $\mbox{su}_2$} 

We are now in a position to introduce a realization of the generators of the
non-deformed Lie algebra  $\mbox{su}_2$ 
in terms of the operators $U_r$ and $H$.  
By using the Schwinger trick~$^{19}$ 
     $$
  j = {1 \over 2} \left(  n_a + n_b  \right), \
  m = {1 \over 2} \left(  n_a - n_b  \right)  \ \Rightarrow \ 
  | n_a ,  n_b ) = |j + m, j-m) \equiv |j , m \rangle
       $$
we can construct a subspace 
$$
{\cal F}_j = \{ |  j , m \rangle : m = -j, -j+1, \cdots, j \}
$$
of the space ${\cal F}$ corresponding to 
     $$
  j = \frac{k-1}{2} \quad {\rm with} \quad k \in {\bf N} \setminus \{ 0,1 \} 
       $$
The possible values of $j$ are thus $j = \frac{1}{2}, 1, \frac{3}{2}, \cdots$.
The value $j = 0$ can be seen to correspond to the limiting situation where 
$k \to \infty$. 

The space  ${\cal F}_j$  of dimension  $2j+1$  is stable 
under $H$ and $U_r$. Indeed, the action of the operators 
$H$ and $U_r$ on ${\cal F}_j$ are given via 
     $$
  H |j , m \rangle = {\sqrt{ (j+m)(j-m+1) }} |j , m \rangle
       $$
and
     $$
  U_r |j , m \rangle = \left[ 1 - \delta (m,j) \right] |j, m+1 \rangle \ + \ 
  \delta(m,j) {\rm exp} 
            \left( + {{\rm i}{{\phi}_r }} \right) |j, -j \rangle 
       $$
In addition, the action on  ${\cal F}_j$ of the adjoint 
$ U_r^{\dagger} $ of  $U_r$ reads 
     $$
  U_r^{\dagger} | j , m   \rangle = \left[ 1-\delta(m,-j) \right] 
                | j , m-1 \rangle \ + \ \delta(m,-j) 
  {\rm exp} \left( - {{\rm i}{{\phi}_r }} \right) 
  | j j \rangle
       $$
(The operator $U_r^{\dagger}$ is defined in terms of the adjoints 
  $a_-^{\dagger}$,
  $a_+^{\dagger}$,
  $b_-^{\dagger}$ and 
  $b_+^{\dagger}$ of  
  $a_-$,
  $a_+$,
  $b_-$ and 
  $b_+$, respectively.) 

It is sometimes useful to use the Dirac notation by
writing 
  \begin{eqnarray*}
H   &=& \sum_{m=-j}^{j  } {\sqrt{ (j+m)(j-m+1) }} |j , m \rangle 
                                                               \langle j, m | \\
U_r           &=& \sum_{m=-j}^{j-1} |j, m+1 \rangle \langle j , m | \ + \ 
  {\rm exp} \left(+{{\rm i}{{\phi}_r }} \right) |j, -j \rangle \langle j, j | \\
U_r^{\dagger} &=& \sum_{m=-j+1}^{j} |j, m-1 \rangle \langle j , m | \ + \
  {\rm exp} \left(-{{\rm i}{{\phi}_r }} \right) |j,  j \rangle \langle j,-j |
  \end{eqnarray*}
It is understood that the three preceding relations 
 are valid as far as the operators $H$, $U_r$ and $U_r^{\dagger}$ act on the
space ${\cal F}_j$. 

We can easily check that the operator $H$ is Hermitean 
and the operator $U_r$ is unitary. As a further property, we have the identity 
     $$
  \left( U_r \right)^{2j+1} = {\rm exp} \left( {\rm i} \phi_r \right) 
       $$  
modulo its action on the space ${\cal F}_j$ 
(the unit matrix of dimension $2j+1$ is supposed to occur in the 
right-hand side of the identity). 
The latter relation reflects the cyclical character of the operator $U_r$
on the space ${\cal F}$. 

Let us introduce the three operators
     $$
  J_+ = H           U_r, \quad 
  J_- = U_r^{\dagger} H, \quad 
  J_3 = {1 \over 2} \left(  N_a - N_b  \right)
       $$  
as fonctions of the generators 
$\left( a_{-}, a_{+}, N_a \right)$ and
$\left( b_{-}, b_{+}, N_b \right)$ of the algebras $A$ and $B$, respectively. 
It is immediate to check that the action on the state $ | j , m \rangle $ 
of the operators $J_{+}$, $J_{-}$ and $J_3$ is given by 
  \begin{eqnarray*}
  J_+ |j, m \rangle &=& {\sqrt{ (j - m)(j + m + 1) }} |j, m + 1 \rangle \\
  J_- |j, m \rangle &=& {\sqrt{ (j + m)(j - m + 1) }} |j, m - 1 \rangle 
  \end{eqnarray*}
and
     $$
  J_3   |j , m \rangle = m |j , m \rangle
       $$
Consequently, we have the commutation relations 
     $$
  \left[ J_3,J_{+} \right] = + J_{+}, \quad
  \left[ J_3,J_{-} \right] = - J_{-}, \quad \left[ J_+,J_- \right] = 2J_3 
       $$
which correspond to the Lie algebra of the  group  $\mbox{SU}_{2}$. 

\section{An Alternative Basis for $\mbox{SU}_{2}$}

The decomposition of the shift operators $J_+$ and 
$J_-$ in terms of $H$ and $U_r$ coincides with the 
polar decomposition introduced  by  L\'evy-Leblond 
in a completely different way.~$^{20}$ 
This is easily seen by taking the matrix elements of
$U_r$ and $H$ and by comparing these elements to the ones of the operators 
$\Upsilon$ and $J_T$ in Ref.~20. This yields $H \equiv J_T$ and, 
by identifying the arbitrary phase $\varphi$ of Ref.~20 
to $\phi_r$, we obtain that $U_r \equiv \Upsilon$. 
In the present paper, we take $\phi_r$ as 
$$
\phi_r = 2 \pi j r 
$$
where $r \in {\bf R}$.

It is easy to prove that the Casimir operator 
     $$
  J^2 = {1 \over 2} \left( J_+J_- +  
                           J_-J_+ \right) + J_3^2 
       $$
or 
     $$
  J^2 = H^2 + J_3^2 - J_3 = U_r ^{\dagger}H^2 U_r + J_3^2 + J_3 
       $$
commutes with $U_r$ for any value of $r$. For  $r$  fixed,
the commuting set $\{ J^2, U_r\}$ provides us with an alternative to the
familiar commuting set $\{ J^2, J_3 \}$ of angular momentum theory. 
The complete set of commuting operators $\{ J^2, U_r\}$ can be easily
diagonalized. This leads to the following result.

{\bf Result}~: The eigenvalues and the common eigenfunctions 
of the operators $U_r$ and $J^2$ are given by
     $$
  U_r | j , \alpha ; r \rangle = 
  \exp \left( - {\rm i} \alpha {2 \pi \over 2 j + 1 } \right)  
      | j , \alpha ; r \rangle, \quad 
  J^2 | j , \alpha ; r \rangle = j(j+1) 
      | j , \alpha ; r \rangle 
       $$
where 
     $$
  |j , \alpha ; r \rangle = {1 \over {\sqrt{2j + 1}}} 
  \sum_{m = -j}^j
  \exp \left( {\rm i} \alpha m {2 \pi \over 2 j + 1 } \right) 
  |j , m \rangle 
       $$
with the range of values 
     $$
  \alpha = - jr, - jr + 1, \cdots, -jr + 2j, \quad 2j \in {\bf N}
       $$
where $r \in {\bf R}$.

The index $\mu$ used in the introduction is here of the form 
$\mu \equiv \alpha;r$. It is to be noted that the label $\alpha$
goes, by step of 1, from $-jr$ to $-jr + 2j$. The inter-basis 
expansion coefficients 
   $$
\langle j m | j \alpha ; r \rangle = {1 \over \sqrt{2 j + 1}} 
  \exp \left( {\rm i} \alpha m {2 \pi \over 2 j + 1 } \right)
     $$
(with 
$m      = -j , -j  + 1, \cdots,        j$ 
and 
$\alpha = -jr, -jr + 1, \cdots, -jr + 2j$) define a unitary
transformation that allows to pass from the well-known 
orthonormal standard basis 
$\{ |j , m \rangle : 2j \in {\bf N}, \ m = - j, - j + 1, \cdots, j \}$
to the orthonormal non-standard basis
$\{ |j , \alpha ; r \rangle : 2j \in {\bf N}, \ 
         \alpha = - jr, - jr + 1, \cdots, -jr + 2j \}$. 
Consequently, the expansion  
     $$
  |j , m \rangle = {1 \over {\sqrt{2j + 1}}} 
  \sum_{\alpha = -jr}^{-jr + 2j}
  \exp \left( - {\rm i} \alpha m {2 \pi \over 2 j + 1 } \right)
    |j , \alpha ; r \rangle  
       $$
with 
     $$
  m = - j, - j + 1, \cdots, j, \quad 2j \in {\bf N}
       $$
renders possible the passage from the non-standard basis to the standard 
basis. The non-standard basis presents some characteristics of both a 
group-subgroup type basis and of a non group-subgroup type basis in the sense 
that 
the label $\alpha ; r$ does not correspond to some irreducible representation 
of a subgroup of $\mbox{SU}_{2}$ and that the subspace 
$\{ |j , \alpha ; r \rangle : \alpha = - jr, - jr + 1, \cdots, -jr + 2j \}$
spans a reducible representation of the cyclic subgroup $C_{2j+1}$ of 
$\mbox{SO}_{3}$.
                           
\section{A New Approach to the 
           Wigner-Racah Algebra of $\mbox{SU}_{2}$}

\subsection{Coupling and recoupling coefficients in the 
             $\{ J^2, U_r \}$ scheme}

For $r$ fixed, the Clebsch-Gordan coefficients (CGc's)
$( j_1 j_2 \alpha_1 \alpha_2 | j \alpha ; r )$ 
in the $\{ J^2, U_r \}$ 
scheme are simple linear combinations of the 
$\mbox{SU}_{2} \supset \mbox{U}_{1}$ CGc's. In fact, we have
   \begin{eqnarray*} 
  \left( j_1 j_2 \alpha_1 \alpha_2 |j \alpha ; r \right) = 
  {1 \over \sqrt{(2j_1 + 1) (2j_2 + 1) (2j   + 1)}} 
  \sum_{m_1=-j_1}^{j_1} 
  \sum_{m_2=-j_2}^{j_2} 
  \sum_{m  =-j  }^{j  }                                        \\ 
  \times q^{\alpha m} q_1^{- \alpha_1 m_1} q_2^{- \alpha_2 m_2} 
  ( j_1 j_2 m_1 m_2 | j m ) 
   \end{eqnarray*}
where
     $$
q   = \exp \left( {\rm i} {2 \pi \over 2 j   + 1 } \right), \quad
q_1 = \exp \left( {\rm i} {2 \pi \over 2 j_1 + 1 } \right), \quad
q_2 = \exp \left( {\rm i} {2 \pi \over 2 j_2 + 1 } \right)
       $$
The CGc's in the $\{ J^2, U_r \}$ scheme 
satisfy the orthonormality relations 
     $$
  \sum_{j \alpha} 
( j_1 j_2 \alpha_1 \alpha_2 | j \alpha ; r )
( j_1 j_2 \alpha_1'\alpha_2'| j \alpha ; r )^*
= \delta (\alpha_1' , \alpha_1)
  \delta (\alpha_2' , \alpha_2)
       $$  
and
     $$
  \sum_{\alpha_1 \alpha_2}
( j_1 j_2 \alpha_1 \alpha_2 | j \alpha ; r )^*
( j_1 j_2 \alpha_1 \alpha_2 | j'\alpha'; r )
= \Delta ( j | j_1 \otimes j_2 ) 
  \delta (j' , j) \delta (\alpha' , \alpha)
       $$  
with $\Delta ( j | j_1 \otimes j_2 ) =1$ or 0 according to as the
familiar Kronecker product $(j_1) \otimes (j_2)$ 
contains or does not contain 
the irreducible representation $(j)$ of $\mbox{SU}_{2}$. Observe 
that both orthonormality relations correspond to a fixed value of the
real number $r$. 

The symmetry properties of the CGc's 
$( j_1 j_2 \alpha_1 \alpha_2 | j \alpha ; r )$
cannot be expressed in a simple way
(except the particular symmetry under the interchange 
$j_1 \alpha_1 \leftrightarrow 
 j_2 \alpha_2$). 
Let us introduce the ${\bar f}_r$ symbol via 
   \begin{eqnarray*} 
  \bar f_r \pmatrix{
  j_1     &j_2     &j_3     \cr
  \alpha_1&\alpha_2&\alpha_3\cr
  } = 
  {1 \over {\sqrt{(2j_1 + 1) (2j_2 + 1) (2j_3 + 1)} } }
  \sum_{m_1 = -j_1}^{j_1} 
  \sum_{m_2 = -j_2}^{j_2} 
  \sum_{m_3 = -j_3}^{j_3}                               \\
  \times q_1^{ - \alpha_1 m_1 } 
         q_2^{ - \alpha_2 m_2 } 
         q_3^{ - \alpha_3 m_3 } 
  \pmatrix{
  j_1&j_2&j_3\cr
  m_1&m_2&m_3\cr
  }
   \end{eqnarray*}
where
$$ 
q_a = \exp \left( {\rm i} {2 \pi \over 2 j_a + 1 } \right) \quad 
\mbox{with} \quad a =1,2,3
$$
The $3 - jm    $ symbol on the right-hand side of the expansion of 
the ${\bar f}_r$ symbol is an ordinary 
Wigner symbol for the group $\mbox{SU}_{2}$ in the 
$\mbox{SU}_{2} \supset \mbox{U}_{1}$ basis. The 
${\bar f}_r$ symbol exhibits the same symmetry properties under 
permutations of its columns as the $3-jm$ Wigner symbol~: Its value 
is multiplied by $(-1)^{j_1 + j_2 + j_3}$ under an odd permutation 
and does not change under an even permutation. It is possible to connect 
the ${\bar f}_r$ symbol to the CGc's in the $\{ J^2, U_r \}$ scheme 
via the introduction of a metric tensor (cf.~Ref.~6). 
The values of ${\bar f}_r$ symbols (as well as the ones of the 
the CGc's of $\mbox{SU}_{2}$ in the $\{ J^2, U_r \}$ scheme)  
are not necessarily real 
numbers. For instance, we have the following property under complex conjugation
     $$
  \bar f_r \pmatrix{
  j_1     &j_2     &j_3     \cr
  \alpha_1&\alpha_2&\alpha_3\cr
  }^* =  (-1)^{j_1 + j_2 + j_3} 
  \bar f_r \pmatrix{
  j_1     &j_2     &j_3     \cr
  \alpha_1&\alpha_2&\alpha_3\cr
  }
       $$
Hence, the value of the ${\bar f}_r$ symbol 
is real if $j_1 + j_2 + j_3$ is 
even and pure imaginary if  $j_1 + j_2 + j_3$ is odd. The behavior of the 
${\bar f}_r$ symbol under complex conjugation is thus 
very different from the one of the ordinary $3-jm$ Wigner symbol. 

The recoupling coefficients 
of the group $\mbox{SU}_{2}$ can be
expressed in terms of coupling coefficients of  $\mbox{SU}_{2}$  in the  
$\{ J^2 , U_r \}$ scheme. 
As an example, the decomposition of  the $6-j$ symbol   as a sum of products of
four ${\bar f}_r$ symbols requires the introduction of six
metric tensors corresponding to the six arguments of the $6 - j$ symbol. 
In addition, the $9-j$ symbol can be expressed 
in terms ${\bar f}_r$ symbols by replacing
the $3-jm$ symbols by ${\bar f}_r$ symbols
in the decomposition of the  $9-j$ symbol  in terms of
a sum of products of six $3-jm$ symbols. The reader may consult Ref.~6 
for general formulas for passing from the $\{ j m \}$ quantization scheme
to the $\{ j \mu \}$ quantization scheme.

\subsection{Wigner-Eckart theorem in the $\{ J^2, U_r \}$ scheme}

From the spherical components $T^{(k)}_m$ (with $k$ fixed and 
$m = -k, -k+1, \cdots, k$) 
of an $\mbox{SU}_{2}$ irreducible tensor operator ${\bf T}^{(k)}$, we define 
the $2 k + 1$ components 
     $$
  T^{(k)}_\alpha (r) = {1 \over {\sqrt{2k+1}}} \sum_{m=-k}^k  
  \exp \left( {\rm i} \alpha m {2 \pi \over 2 k + 1 } \right) T^{(k)}_m 
       $$
with $\alpha = -kr, -kr + 1, \cdots, -kr + 2k$. 
In the $\{ J^2, U_r \}$ scheme, the Wigner-Eckart theorem reads
    $$
 \langle \tau_1 j_1 \alpha_1 ; r | T^{(k)}_{\alpha}(r) | 
         \tau_2 j_2 \alpha_2 ; r \rangle = \left( 
         \tau_1 j_1             || T^{(k)}               || 
         \tau_2 j_2 \right) 
 f_r\pmatrix{
 j_1     &j_2     &k     \cr
 \alpha_1&\alpha_2&\alpha\cr
 }
      $$
where
     $$
  f_r\pmatrix{
  j_1     &j_2     &j_3     \cr
  \alpha_1&\alpha_2&\alpha_3\cr
  }
  = (-1)^{2j_3} {1 \over {\sqrt{2j_1+1}}} 
  \left( j_2 j_3 \alpha_2 \alpha_3 | j_1 \alpha_1 ; r \right)^*
       $$
The quantity $\left( \tau_1 j_1 || T^{(k)} || \tau_2 j_2 \right)$ 
denotes an ordinary reduced matrix
element. Such an element is basis-independent. Therefore, it does not depend 
on the labels $\alpha_1$, $\alpha_2$ and $\alpha$. On the contrary, 
the $f_r$ coefficient 
depends on the labels $\alpha_1$, $\alpha_2$ and $\alpha$. 
The ${f}_r$ symbol (which is less symmetrical than the ${\bar f}_r$ symbol) 
can be related to the ${\bar f}_r$ symbol by using a metric tensor.

\section{Conclusions}

The main results presented in this lecture are~: (i) The
non-deformed Lie algebra $\mbox{su}_{2}$ may be constructed 
from two commuting 
$q$-deformed oscillator algebras with $q$ being a root of unity~; the latter  
oscillator algebras are associated to (troncated) harmonic oscillators having a
finite number of eigenstates. (ii) This construction leads to the polar
decomposition of the generators $J_+$ and 
                                $J_-$ of $\mbox{SU}_{2}$ 
originally introduced by L\'evy-Leblond.~$^{20}$ (iii) 
The familiar $\{ J^2, J_3 \}$ 
scheme with the standard spherical basis 
$\{ |j , m \rangle : 2j \in {\bf N}, \ m = -j, -j+1, \cdots, j\}$, 
corresponding to
the canonical chain of groups $\mbox{SU}_{2} \supset \mbox{U}_1$, is thus 
replaced by the 
$\{ J^2, U_r \}$ scheme with another basis, 
namely, the non-standard basis    
$\{ |j , \alpha ; r \rangle : 2j \in {\bf N}, 
       \ \alpha = -jr, -jr+1, \cdots, -jr + 2j\}$. 
(iv) The Wigner-Racah algebra of  $\mbox{SU}_{2}$  may be developed in the 
$\{ J^2, U_r \}$ scheme.  These various results shall be further developed 
in a forthcoming paper 
and contact with Quantum Mechanics on a finite Hilbert space, 
as developed by Vourdas,~$^{21}$ shall be established.

\section*{Acknowledgments}

The quon approach to $\mbox{su}_{2}$ 
was developed in
collaboration with Dr.~M.~Daoud (see Ref.~17). 
The author thanks Dr.~M.~Daoud for many discussions. He is grateful to the 
Organizing Committee of the 5th International School of Theoretical Physics 
SSPCM98, and more specifically Prof.~T.~Lulek, for inviting 
him to deliver this lecture. He is also indebted to Profs.~M.~Bo\.zejko 
and A.~Vourdas for interesting comments.


\section*{References}

\begin{thebibliography}{99}

 \bibitem{1-EPW} E.P. Wigner, 
 in {\em Quantum Theory of Angular Momemtum}, 
 eds. L.C. Biedenharn and H. van Dam
 (Academic Press, New York, 1965). 

 \bibitem{2-GR} G. Racah, 
 in {\em Quantum Theory of Angular Momemtum}, 
 eds. L.C. Biedenharn and H. van Dam
 (Academic Press, New York, 1965). 

 \bibitem{3-WL} W. Low, 
 in {\em Quantum Electronics I and II} (Columbia University Press, 1960 and
 1961)~;  
 in {\em Spectroscopic and Group Theoretical Methods in Physics}, 
 eds. F. Bloch, S.G. Cohen, A. de-Shalit, S. Sambursky and I. Talmi 
 (North-Holland, Amsterdam, 1968). 

 \bibitem{4-LR}  W. Low and G. Rosengarten, 
                            {\em J. Molec. Spectrosc.}   {\bf 12}, 319  (1964). 

 \bibitem{5-MF}  M. Flato,  {\em J. Molec. Spectrosc.}   {\bf 17}, 300  (1965). 

\bibitem{6-KIB} M. Kibler, 
                        {\em J. Molec. Spectrosc.} {\bf 26}, 111 (1968)~; 
                        {\em Int. J. Quantum Chem.} {\bf 3}, 795 (1969)~; 
                        {\em C.~R. Acad. Sci. (Paris) B} {\bf 268}, 1221 (1969).

 \bibitem{7} M.R. Kibler, {\em J. Math. Phys.}         {\bf 17}, 855  (1976)~; 
                          {\em J. Molec. Spectrosc.}   {\bf 62}, 247  (1976)~; 
                          {\em J. Phys. A: Math. Gen.} {\bf 10}, 2041 (1977).

 \bibitem{8-L} T. Lulek, {\em Acta Phys. Polon. A} {\bf 43}, 705 (1973) and 
                                                   {\bf 48}, 501 (1975). 

 \bibitem{9-L} B. Lulek,    T. Lulek and B. Szczepaniak, 
 {\em Acta Phys. Polon. A} {\bf 54}, 545 (1978)~; 
               B. Lulek and T. Lulek, 
 {\em Acta Phys. Polon. A} {\bf 54}, 561 (1978). 

 \bibitem{10-L} B. Lulek, 
 {\em Acta Phys. Polon. A} {\bf 55}, 165 (1979). 

\bibitem{11-MB} J. Moret-Bailly, 
 {\em J. Mol. Spectrosc.} {\bf 15}, 344 (1965)~; 
 J.P. Champion, G. Pierre, F. Michelot and J. Moret-Bailly, 
 {\em Can. J. Phys.} {\bf 55}, 512 (1977). 

\bibitem{12-PW} J. Patera and P. Winternitz, {\em J. Math. Phys.} {\bf 14},
1130 (1973)~; {\em J. Chem. Phys.} {\bf 65}, 2725 (1976).  

\bibitem{13-Mich} L. Michel, in {\em Group Theoretical Methods in Physics}, 
eds. R.T. Sharp and B. Kolman (Academic Press, New York, 1977). 

\bibitem{14-Ari19} M.~Arik and D.D.~Coon, 
{\em J.~Math.~Phys.} {\bf 17}, 524 (1976). 

\bibitem{15-26} M. Daoud, Y. Hassouni and M. Kibler, 
in {\em Symmetries in Science X},
eds. B. Gruber and M. Ramek (Plenum Press, New York, 1998)~; 
{\em Yad. Fiz.} (to appear). 

\bibitem{16-01} J.M. Leinaas and J. Myrheim, {\em Nuovo Cimento B} {\bf 37},  
1 (1977).       G.A. Goldin, R. Menikoff and D.H. Sharp, {\em J. Math. Phys.} 
{\bf 21}, 650  (1980) and 
{\bf 22}, 1664 (1981). 

 \bibitem{17-KIDA} M. Kibler and M. Daoud, 
 {\em Turkish J. Phys.} (to appear). 

\bibitem{18-25} D.B. Fairlie, P. Fletcher and C.K. Zachos, {\em J. Math. Phys.} 
{\bf 31}, 1088 (1990). 

 \bibitem{19-JS} J. Schwinger,  
 in {\em Quantum Theory of Angular Momemtum}, 
 eds. L.C. Biedenharn and H. van Dam
 (Academic Press, New York, 1965). See also 
              M. Kibler and G. Grenet, 
 {\em J. Math. Phys.} {\bf 21}, 422 (1980). 

\bibitem{20-28} J.-M. L\'evy-Leblond, {\em Rev. Mex. F\'\i sica} {\bf 22},  
15 (1973). 

\bibitem{21} A. Vourdas, {\em Phys. Rev. A} {\bf 41}, 1653 (1990) and 
                                            {\bf 43}, 1564 (1991)~; 
             A. Vourdas and C. Bendjaballah, 
                         {\em Phys. Rev. A} {\bf 47}, 3523 (1993)~;
             A. Vourdas, {\em J. Phys. A: Math. Gen.} {\bf 29}, 4275 (1996). 
 
\end{thebibliography}
\end{document}